\documentclass[aps,onecolumn,superscriptaddress,groupedaddress,nofootinbib]{revtex4}

\usepackage{graphicx}
\usepackage{dcolumn}
\usepackage{bm}
\usepackage{mathrsfs}
\usepackage{hyperref}
\usepackage{amsmath}
\usepackage{amssymb}
\usepackage{bm}
\usepackage{color}
\usepackage{float}
\usepackage{dcolumn}
\usepackage{multirow}
\usepackage{changepage}
\usepackage{enumerate}
\hypersetup{pdftex,colorlinks=true,linkcolor=blue,citecolor=red,menucolor=black,urlcolor=blue,filecolor=blue}

\newcommand{\mev}{\textrm{ MeV}}

\begin{document}
\title{The $\bar{B}^0 \to D^{*+}  \bar{D}^{*0} K^-  $  reaction to detect the $I=0, J^P=1^+$ partner of the $X_0(2866)$ }
\date{\today}

\author{L.~R. Dai}
\email{dailianrong@zjhu.edu.cn}
\affiliation{School of Science, Huzhou University, Huzhou 313000, Zhejiang, China}
\affiliation{Departamento de F\'{\i}sica Te\'orica and IFIC, Centro Mixto Universidad de Valencia-CSIC Institutos de Investigaci\'on de Paterna, Aptdo.22085, 46071 Valencia, Spain}

\author{R.~Molina}
\email{Raquel.Molina@ific.uv.es}
\affiliation{Departamento de F\'{\i}sica Te\'orica and IFIC, Centro Mixto Universidad de
Valencia-CSIC Institutos de Investigaci\'on de Paterna, Aptdo.22085,
46071 Valencia, Spain}

\author{E.~Oset}
\email{oset@ific.uv.es}
\affiliation{Departamento de F\'{\i}sica Te\'orica and IFIC, Centro Mixto Universidad de
Valencia-CSIC Institutos de Investigaci\'on de Paterna, Aptdo.22085,
46071 Valencia, Spain}

\begin{abstract}
We have chosen the  $\bar{B}^0 \to D^{*+}  \bar{D}^{*0} K^-$  reaction  in order to observe the $I=0, J^P=1^+$ ($R_1$) partner state of the
$X_0(2866)$  stemming from the $D^*\bar{K}^*$ molecular picture. The reaction proceeds via external emission  in the most favored Cabibbo decay
mode and one observes the  $R_1$ state as a very strong peak versus the background in the $D^{*+} K^-$ spectrum. The branching ratio for $R_1$ production
in this reaction is estimated of the order of $4\times 10^{-3}$. The method used, applied to the $B^+ \to D^- D^+ K^+$ reaction, produces a ratio of signal to
background  in the  $D^- K^+$ spectrum in very good agreement with the LHCb experiment that observed the $X_0(2866)$.

\end{abstract}

\maketitle

\section{Introduction}
The discovery of the  $X_0(2866)$ and $X_1(1900)$ states by the LHCb collaboration \cite{lhcb1,lhcb2} was a turning point in hadron spectroscopy, offering the first clear example of an exotic meson with two open flavor quarks of type $cs\bar{u}\bar{d}$, which cannot be cast in terms of
the conventional $q\bar{q}$ structure of the mesons. The recent discovery of the $T_{cc}$ state, again by the LHCb
collaboration \cite{tcc1,tcc2}, has come to boost this discovery  and make these exotic states part of the our ordinary meson zoo.

Three main lines of research are followed to interpret  the $X_0(2866)$. Its structure  as a compact tetraquark state is discussed  in
\cite{weiwang,karliner,oka,segovia}.  The sum rules technique is used in \cite{zhang,zigang,hxchen1,hxchen2,narison} and the molecular picture
as an $I=0, J^P=1^+$ $D^*\bar{K}^*$ molecular state is pursued in \cite{xie,gengx,hu,meng,jhe,slzhu}. Some of the sum rule studies  go deeper
into details and conclude  that the state is of $D^*\bar{K}^*$ molecular nature \cite{hxchen1,hxchen2,narison}. On the other hand,
in a detailed quark  model calculation in \cite{qifang}, the compact tetraquark picture is disfavored.

There are other works suggesting a triangle singularity \cite{xiets} and in \cite{burns,swanson} a discussion is done about possible structures
as a molecular cusp effect or a consequence of analytical properties of triangle diagrams. A triangle mechanism is also used in \cite{qifang2}, using empirical information on $D^-\to D_s^{*-}D^0$ decay followed by $D^{*-}_s\to D^-\bar{K}^0$ and fusion of $D^0\bar{K}^0\to X(2866)$.

The molecular picture is appealing and strongly supported in the works mentioned above. One point in its favor stems from the fact that a prediction, prior to its observation  by the LHCb collaboration, was done in \cite{branz} in remarkable agreement with the experimental information. Indeed, a $D^*\bar{K}^*$ molecule with $I=0$ and  $J^P=0^+$ was predicted  in  \cite{branz} with mass $2848\mev$ and width
between $23-59\mev$.  This is to be contrasted with the experimental result of $M_{X_0}=2866 \pm 7\mev$, $\Gamma_{X_0}=57.2\pm 12.9\mev$.
The width of the $X_0(2866)$  comes in \cite{branz} from the decay to the $D \bar{K}$ channel, where it was observed.

Interestingly, in  \cite{branz} two more states with $I=0$  and $J^P=1^+, 2^+$ were predicted, which are not yet observed. Such states also
appear in other studies of the molecular $D^*\bar{K}^*$ picture as \cite{meng}. An update of the work of \cite{branz}, fine tuning the parameter that regularizes the  $D^*\bar{K}^*$ loops to obtain the precise mass  and width observed in the experiment, has been done in \cite{raquelx}, and then,  using the same parameters, more precise predictions are done for the  $J^P=1^+, 2^+$ states. The information on these states is given in Table \ref{tab:1}.
\begin{table}[t]
\renewcommand{\arraystretch}{1.2}
\begin{center}
\caption{$D^*\bar{K}^*$ states obtained from Ref. \cite{raquelx} including the width to the $D\bar{K}$ and $D^*\bar{K}$ channels \cite{raquelx}.}
 \begin{tabular}{cccccc}
 \hline
 ~~~ $I [J^P]$ ~~~&~~~ $M~[\mathrm{MeV}]$ ~~~ & ~~~$\Gamma~[\mathrm{MeV}]$ ~~~ &~~~ Coupled channels~~~ &~~~ $g_{R,D^*\bar{K}^*}$ $[\mathrm{MeV}]$ &~~~ state~~~\\
  \hline
  $0 [2^+]$& $2775$ & $38$ & $D^*\bar{K}^*$ & $16536$ &?\\
  $0 [1^+]$& $2861$ & $20$ & $D^*\bar{K}^*$ & $12056$ & ?\\
  $0 [0^+]$&$2866$ & $57$& $D^*\bar{K}^*$ & $11276$ & $X_0(2866)$\\
  \hline
 \end{tabular}
\end{center}
\label{tab:1}
\end{table}

The purpose of the present paper is  to suggest a reaction where the $J^P=1^+$  state  can be observed. The reaction is
The $\bar{B}^0 \to D^{*+}  \bar{D}^{*0} K^-$  looking at the $D^{*+} K^-$  invariant mass distribution. The reasons that lead us to
this particular reaction are the following:
\begin{itemize}
  \item[a)] In  \cite{babar} a long list of reactions was measured of the type $B \to D^{(*)}  \bar{D}^{(*)} K$ and classified along its
  topological decay mode as external emission, internal emission  or mixture of the two. Out of this, the $B^0 \to D^{*-} D^{*0} K^+$ reaction
  is favored  because  it proceeds via external emission (favoring the decay), has the largest branching fraction, $1.06\%$, and can produce
  the $D^{*-} K^+$ in $I=0$.
  \item[b)]  The $D^{*-} K^+$ decay mode is the one where the $J^P=1^+$ state can decay \cite{raquelx}. Indeed, this state cannot decay to
  $\bar{D}K$, where the $X_0(2866)$ was found, because it needs $L=1$ and violates parity. On the other hand, the $J^P=0^+$ state cannot decay
  to $\bar{D}^* K$ for the same reasons.
\end{itemize}
We are then in a situation where the observation of a peak in the $D^{*-} K^+$ invariant mass distribution could clearly be associated to a
$1^+$ state (we shall discuss the $2^+$ possibility later). We show the technical details in the next section.

\section{FORMALISM}

\begin{figure}[h]
\centering
\includegraphics[scale=0.85]{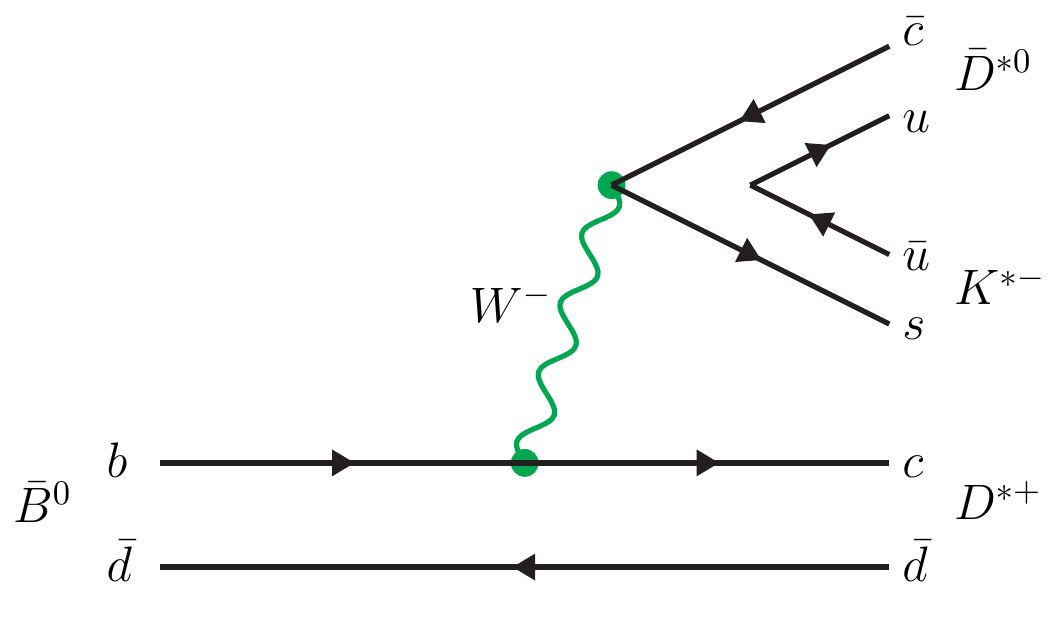}
\caption{Diagrammatic decay of the $\bar{B}^0 \to  \bar{D}^{*0} D^{*+}  K^{*-} $ at the quark level.}
\label{fig:1}
\end{figure}

For convenience we study the charge conjugate reaction and choose the  $\bar{B}^0 \to D^{*+}  \bar{D}^{*0} K^-$  one. The $D^{*+} K^-$  is in our case the decay product of $D^*\bar{K}^*$ in $I=0$ and $J^P=1^+$. Hence, the primary reaction that we have is $\bar{B}^0 \to D^{*+}  \bar{D}^{*0} K^{*-}$. This reaction proceeds via external emission, as depicted in Fig.~\ref{fig:1}.

\begin{figure}[h]
\centering
\includegraphics[scale=0.8]{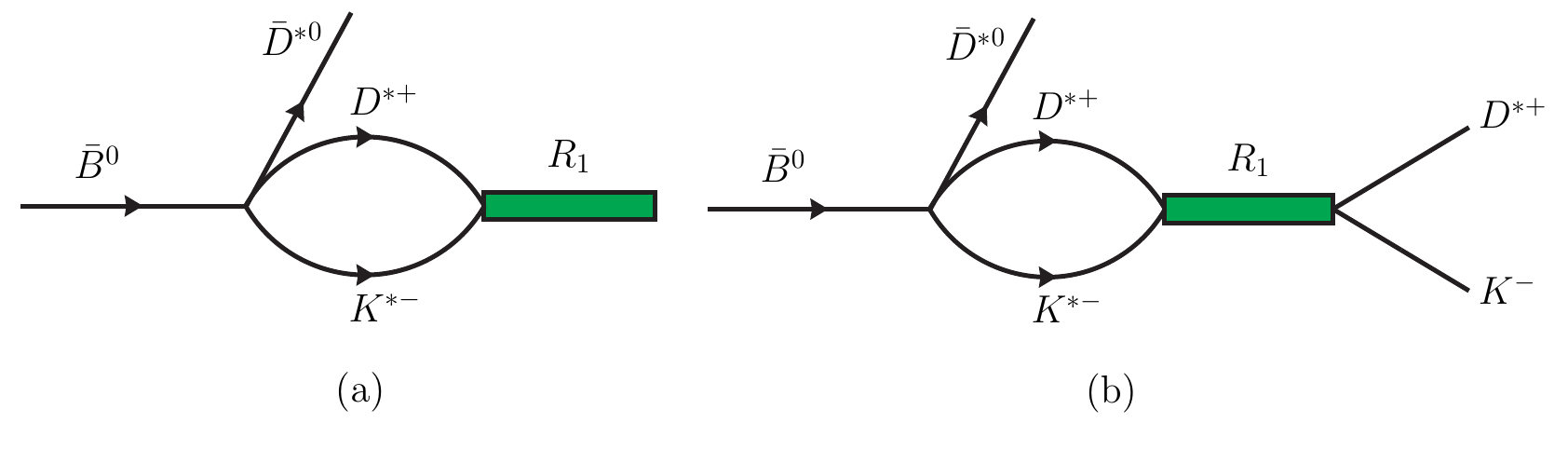}
\caption{(a) Rescattering of $D^{*+}  K^{*-}$ to give the resonance $R_1$ of $I=0, J^P=1^+$;
(b) Further decay of $R_1$ to  $D^{*+} K^-$.}
\label{fig:2}
\end{figure}

The hadronization of the $\bar{c}s$ component with a $u\bar{u}$ pair gives us the hadron state $\bar{D}^{*0} D^{*+}  K^{*-}$. The next step is shown in Fig.~\ref{fig:2}  where the $D^{*+}  K^{*-}$ interacts to give the $I=0, J^P=1^+$ state ($R_1$).

With the isospin multiplets  $(D^{*+},-D^{*0})$, $(\bar{D}^{*0},D^{*-})$,  $(K^{*+},K^{*0})$,  $(\bar{K}^{*0},-K^{*-})$, we have the $I=0$ state
\begin{eqnarray}
|D^* \bar{K}^*\rangle = -\frac{1}{\sqrt{2}}(D^{*+}K^{*-}+D^{*0}\bar{K}^{*0} ) \,.
\end{eqnarray}

On the other hand, we need to have the structure of the decay vertex, $\bar{B}^0 \to  \bar{D}^{*0} D^{*+}  K^{*-} $. Since total spin is conserved in the weak decay, we need to construct a scalar with the three polarization vectors of the vector mesons. Assuming $s$-wave dominance in the decay, this vertex is given by
\begin{eqnarray}
-i\, t = -i\, C\, {\bm \epsilon^{(1)}}\cdot ({\bm \epsilon^{(2)}} \times {\bm \epsilon^{(3)}})= -i\, C\, \epsilon_{ijk}\epsilon^{(1)}_i
\epsilon^{(2)}_j  \epsilon^{(3)}_k  \nonumber
\end{eqnarray}
where the indices $1,2,3$ apply to the $\bar{D}^{*0}$,$ D^{*+}$ and $K^{*-}$ respectively, and $C$ is an unknown constant.

We need another ingredient, which is the spin projection operator for the $1^+$ state. The spin projectors are given in \cite{raquelvec} for
the $VV \to VV$ amplitude and more convenient for us, factorizing the two vertices in $VV \to R \to VV$, in \cite{moliliang}.
The vertex $R_l \to VV$ for $l=0,1,2$ the total spin of the two vectors, are
\begin{eqnarray}
V^{(0)}&=& \frac{1}{3} \epsilon^{(2)}_l \epsilon^{(3)}_l \delta_{ij} \nonumber \\
V^{(1)}&=& \frac{1}{2} \big(\epsilon^{(2)}_i \epsilon^{(3)}_j -\epsilon^{(2)}_j \epsilon^{(3)}_i \big)     \nonumber  \\
V^{(2)}&=&   \frac{1}{2} \big(\epsilon^{(2)}_i \epsilon^{(3)}_j +\epsilon^{(2)}_j \epsilon^{(3)}_i \big)- \frac{1}{3} \epsilon^{(2)}_l \epsilon^{(3)}_l \delta_{ij}
\end{eqnarray}

Considering that in the propagators  in  Fig.~\ref{fig:2} (a) we have the sum over the polarization of the $D^{*+}$ and $K^{*-}$ vectors,
$\sum_{pol}\epsilon^{(r)}_i \epsilon^{(r)}_j=\delta_{ij} ~(r=2,3)$, the amplitude in Fig.~\ref{fig:2} (a) is given by
\begin{eqnarray}
-i\, \tilde{t} = -i\, C\, \epsilon^{(1)}_i \epsilon_{i i' j'} G_{D^*\bar{K}^*} (M_{\rm inv})  \frac{-1}{\sqrt{2}} g_{R,D^*\bar{K}^*} \nonumber
\end{eqnarray}
where $i', j'$ are the indices of the polarization vectors of the $D^*,\bar{K}^*$ vectors forming the $R_1$ molecule and
$g_{R,D^*\bar{K}^*}$  is the coupling of the resonance $R_1$ to the $D^*\bar{K}^*$ $(I=0)$ state.  $G_{D^*\bar{K}^*} (M_{\rm inv})$
is the loop function of the $D^*,\bar{K}^*$  propagators  and we use the same one as used in \cite{raquelx} to obtain the mass of the
$X_0(2866)$  state, and $M_{\rm inv}$ is the invariant mass of the $R_1$.

The sum of the  contribution of the three different polarization  states of $R_1$ is obtained  summing over  $i', j'$ in $|t|^2$ and
we obtain (summing also over the polarizations of the $\bar{D}^{*0}$ state)
\begin{eqnarray}
\sum_{pol} \epsilon^{(1)}_i \epsilon_{i i' j'}   \epsilon^{(1)}_m \epsilon_{m i' j'} =  \epsilon_{i i' j'}  \epsilon_{i i' j'}
=\delta_{i i'}\delta_{j j'} -\delta_{i' j'}\delta_{j' i'} =9-3=6   \nonumber
\end{eqnarray}
Hence,
\begin{eqnarray}
\sum_{pol}|t|^2 =\frac{6}{2} C^2 |g_{R,D^*\bar{K}^*}|^2 |G_{D^*\bar{K}^*} (M_{\rm inv})|^2 \,.
\label{eq:t2}
\end{eqnarray}

We can already see that with the ${\bm \epsilon^{(2)}} \times {\bm \epsilon^{(3)}}$ operator that couples two $S=1$ vectors to $J=1$,
it is clear that we cannot produce the $0^+$ and  $2^+$ states with this reaction.

 One last step is needed if we want to evaluate the $t'$ matrix for the process of Fig.~\ref{fig:2} (b) which incorporates the decay of the resonance  to $D^{*+} K^-$. The  $|t|^2$  matrix in  Eq.~\eqref{eq:t2} will now incorporate the  $R_1$ propagator and
 an effective coupling $g_{R,D^*\bar{K}^*}$ for the decay of $R_1 \to D^*\bar{K}^*$  in $I=0$. Once  again we obtain a factor $\frac{-1}{\sqrt{2}}$ by projecting the $I=0$ state into the $D^{*+} K^-$ component and we find
\begin{eqnarray}
\sum_{pol}|t'|^2 &=&\frac{6}{4} C^2 |g_{R_1,D^*\bar{K}^*}|^2 |G_{D^*\bar{K}^*} (M_{\rm inv})|^2   \nonumber\\
&\times&|g_{R_1,D^* \bar{K}}|^2 |\frac{1}{M^2_{\rm inv} (R_1)-M^2_{R_1}+i M_{R_1} \Gamma_{R_1}}|^2\label{eq:tp}
\end{eqnarray}
with $M^2_{\rm inv}=(P_{D^{*+}}+P_{K^-})^2$ and $M_{R_1}$, $\Gamma_{R_1}$, $g_{R_1,D^*\bar{K}^*}$ given in Table \ref{tab:1}.
The effective $|g_{R_1,D^* \bar{K}}|^2$   coupling is obtained from the $R_1 \to D^* \bar{K}$ width
\begin{eqnarray}
\Gamma_{R_1}=\frac{1}{8\pi} \frac{1}{M^2_{R_1}} |g_{R_1,D^* \bar{K}}|^2 q_{\bar{K}}\,;\qquad q_{\bar{K}}=\frac{\lambda^{1/2}(M^2_{R_1},m^2_{D^*},m^2_{\bar{K}})}{2 M_{R_1}}
\end{eqnarray}

The  $D^{*+} K^-  $ invariant mass distribution is now given by
\begin{eqnarray}
\frac{d\Gamma}{dM_{\rm inv} (D^{*+}K^-)}=\frac{1}{(2\pi)^3} \frac{1}{4M^2_{\bar{B}^0}} p_{\bar{D}^{*0}} \tilde{p}_{K^-} \sum |t'|^2
\label{eq:t2}
\end{eqnarray}
where
\begin{eqnarray}
 \tilde{p}_{K^-} =\frac{\lambda^{1/2}(M^2_{\rm inv}(D^{*+} K^-),m^2_{D^*},m^2_{\bar{K}})}{2 M_{\rm inv}(D^{*+} K^-)} \,, \quad
 p_{\bar{D}^{*0}}=\frac{\lambda^{1/2}(M^2_{\bar{B}^0},m^2_{\bar{D}^{*0}},M^2_{\rm inv}(D^{*+} K^-)}{2 M_{\bar{B}^0}}
\end{eqnarray}

We have an unknown constant $C$ which we would like to get rid of.  For that purpose we compare the signal obtained for the
 $R_1$ state with the background that we would expect for the $\bar{B}^0 \to  \bar{D}^{*0} D^{*+}  K^{-}$ reaction. For this reaction we will assume that the  amplitude is given   in $s$-wave by
\begin{eqnarray}
-i t = -i C {\bm \epsilon (D^{*0} ) } \cdot {\bm \epsilon (D^{*+} ) }
\label{eq:ano}
\end{eqnarray}
and we assume that $C$ is the same as before. Actually the topology of this reaction is identical to the one of Fig.~\ref{fig:1}  substituting the $K^{*-} $ by a $K^-$.
A detailed study is done in \cite{vectorpseudo} of the nonleptonic weak decay of heavy hadrons into pseudoscalar or
vectors,  and for a same topology the differences are Racah algebra coefficients of the same order of magnitude. Hence, Eq.~\eqref{eq:ano} is
a reasonable assumption. In the next section we will further test this  assumption in the decay of the $R_0$ ($ I=0, J^P=0^+$ state, the
$X_0(2866)$), confirming its fairness.

Going through the same steps as before one easily finds now that
\begin{eqnarray}
\frac{d\Gamma_{\rm bac}}{dM_{\rm inv} (D^{*+}K^-)}=\frac{1}{(2\pi)^3} \frac{1}{4M^2_{\bar{B}^0}} p_{\bar{D}^{*0}} \tilde{p}_{\bar{K}} \,3\, C^2
\label{eq:t2b}
\end{eqnarray}
and we can compare the results of Eq.~\eqref{eq:t2}   with those of the background for the same reaction in Eq.~\eqref{eq:t2b}.

\begin{figure}[h]
\centering
\includegraphics[scale=0.75]{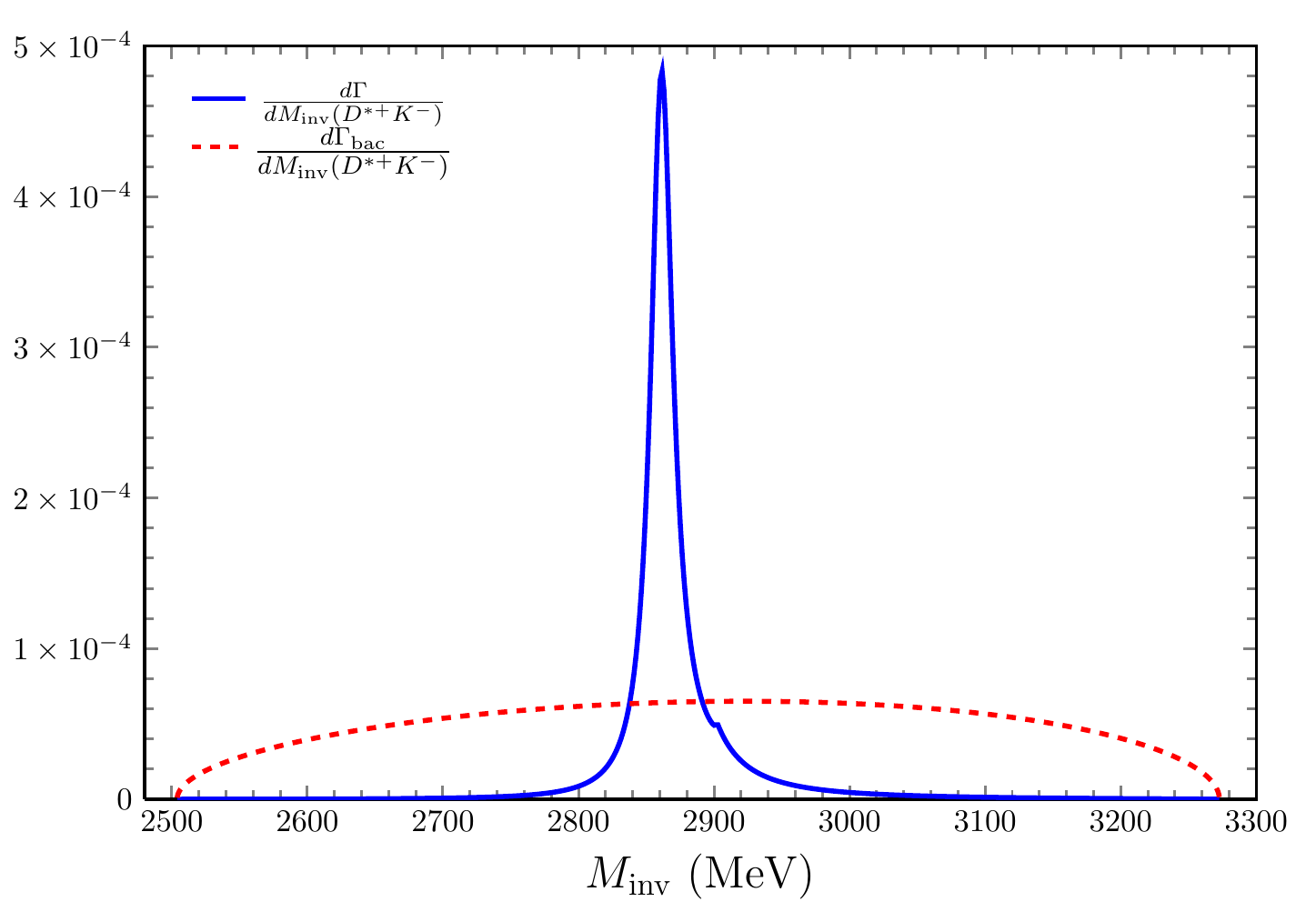}
\caption{$\frac{d\Gamma}{dM_{\rm inv}}$ for the $R_1$ production versus the background, $\frac{d\Gamma_\mathrm{bac}}{dM_{\rm inv}}$, in the $\bar{B}^0 \to  \bar{D}^{*0} D^{*+}  K^{*-}$ reaction in a global arbitrary normalization. $M_{\rm inv}$ is the invariant mass of $D^{*+} K^-$. }
\label{fig:dgR1}
\end{figure}

In Fig.~\ref{fig:dgR1}   we show the results if $\frac{d\Gamma_{\rm bac}}{dM_{\rm inv}}$ for the signal of the $R_1$ resonance compared with the background.

As we  can see,  the peak stands clearly over the background and should be easily observed.  We can also evaluate the branching ratio for $R_1$ production and decay to  $D^{*+} K^-$ in this reaction by integrating over $M_{\rm inv}$ and we find
\begin{eqnarray}
\frac{\int\frac{d\Gamma}{dM_{\rm inv}}}{\int\frac{d\Gamma_{\rm bac}}{dM_{\rm inv}}} = 0.4\,; \qquad {{\cal B}_R} (R_1; R_1 \to D^{*+} K^-)=4.24 \times 10^{-3}
\end{eqnarray}
where in the last step to evaluate the  branching ratio  we have used the  branching ratio of Babar for $\bar{B}^0 \to  \bar{D}^{*0} D^{*+}  K^{*-}$ of  $1.06 \%$ \cite{babar}.  The value obtained is large compared to the typical weak decays of the $B$.  We can even indulge in  a factor of two or three uncertainty in this rate and the peak as well as the branching ratio would still be
 sizeable and easily visible.  But let us  see what the method used gives for the $X_0(2866)$ production.

\section{Results for the $B^- \to D^- D^+ K^+$ reaction with $X_0(2866)$  production}
This reaction proceeds via internal emission \cite{babar} and one expects a smaller  branching ratio than in the former case \cite{babar}.
In Ref. \cite{lhcb1} (Fig. 3 left of that reference) one can see a sharp peak for the $X_0(2866)$ (we call it now $R_0$) in the $D^- K^+$ spectrum,  with strength a few times larger than the background, qualitatively similar to what we have obtained in Fig.~\ref{fig:dgR1}.
We repeat the procedure of the former section with the charge conjugate reaction  $B^- \to D^- D^+ K^+$ and look at the $D^+ K^-$ invariant mass. We start from the $B^- \to D^- D^{*+} K^{*-}$ reaction with a vertex
\begin{eqnarray}
-i t = -i C' {\bm \epsilon (D^{*+} ) } \cdot {\bm \epsilon (K^{*-} ) }  \nonumber
\end{eqnarray}
and the $B^- \to D^- D^+ K^-$  reaction  with a vertex
\begin{eqnarray}
-i \tilde{t}' = -i C'  \nonumber
\end{eqnarray}
as we have done before with the same constant $C'$. It is easy to redo the calculations and we find now
\begin{eqnarray}
\frac{d\Gamma'}{dM_{\rm inv} (D^{+}K^-)}=\frac{1}{(2\pi)^3} \frac{1}{4M^2_{\bar{B}^-}} p_{D^-} \tilde{p}_{K^-} \sum|t'|^2
\end{eqnarray}
where $t'$ includes now the decay part of the resonance in analogy to Eq.~(\ref{eq:tp}),
\begin{eqnarray}
\sum|t'|^2 &=&\frac{3}{4} C'^2 |G_{D^* \bar{K}^*} (M_{\rm inv}(D^+K^-)) |^2  \, |g_{R_0,D^* \bar{K}^*}|^2  \nonumber\\
&\times&  |\frac{1}{M^2_{\rm inv} (D^+ K^-)-M^2_{R_0}+i M_{R_0} \Gamma_{R_0}}|^2 \, |g_{R_0,D \bar{K}}|^2
\end{eqnarray}
with
\begin{eqnarray}
 p_{\bar{D}^{-}}=\frac{\lambda^{1/2}(M^2_{\bar{B}^-},m^2_{\bar{D}^{-}},M^2_{\rm inv}(D^{+} K^-)}{2 M_{\bar{B}^-}} \,, \quad
  \tilde{p}_{K^-} =\frac{\lambda^{1/2}(M^2_{\rm inv}(D^{+} K^-),m^2_{D^+},m^2_{K^-})}{2 M_{\rm inv}(D^{+} K^-)}
\end{eqnarray}
and $g_{R_0,D^* \bar{K}^*}$ given in Table \ref{tab:1}. The effective $|g_{R_0,D \bar{K}}|^2 $ is obtained from
\begin{eqnarray}
\Gamma_{R_0}=\frac{1}{8\pi} \frac{1}{M^2_{R_0}} |g_{R_0,D \bar{K}}|^2 q_{\bar{K}}\,;\qquad q_{\bar{K}}=\frac{\lambda^{1/2}(M^2_{R_0},m^2_{D^+},m^2_{\bar{K}})}{2 M_{R_0}}
\end{eqnarray}

For the background we find
\begin{eqnarray}
\frac{d\Gamma'_{\rm bac}}{dM_{\rm inv} (D^{+}K^-)}=\frac{1}{(2\pi)^3} \frac{1}{4M^2_{\bar{B}^-}} p_{D^{-}} \tilde{p}_{K^-} \, C^2
\end{eqnarray}

\begin{figure}[h]
\centering
\includegraphics[scale=0.75]{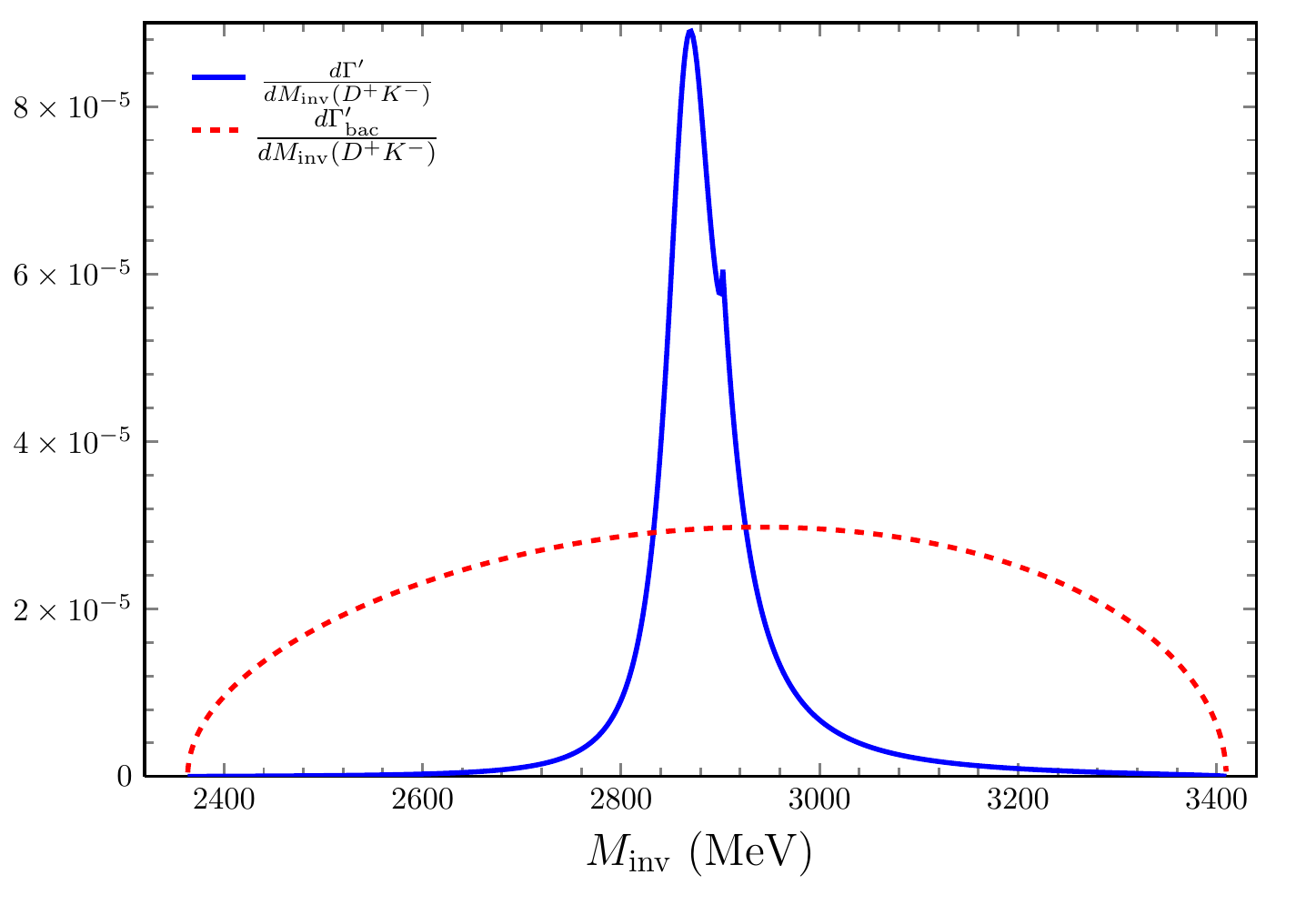}
\caption{$\frac{d\Gamma'}{dM_{\rm inv}}$ for the $R_0$ production versus the background, $\frac{d\Gamma_\mathrm{bac}}{dM_{\rm inv}}$, in the $B^- \to  D^- D^+  K^-$ reaction in a global arbitrary normalization. $M_{\rm inv}$ is the invariant mass of $D^+ K^-$. }
\label{fig:dgR0}
\end{figure}

In Fig.~\ref{fig:dgR0}   we show now the results of $\frac{d\Gamma'}{dM_{\rm inv}}$ and $\frac{d\Gamma'_{\rm bac}}{dM_{\rm inv}}$
 comparing the signal with the background. We observe a structure very similar to the experiment in Fig. 3 left of Ref. \cite{lhcb1}. The
 strength at the peak versus the background is about a factor of $3$ versus a factor of about $3.6$ in the experiment. The success
 of this test gives further strength to the estimates made in the former section.

\section{Conclusions}

We have chosen a reaction,   $\bar{B}^0 \to D^{*+}  \bar{D}^{*0} K^-$ reaction, as a tool to observe the $I=0, J^P=1^+$  state ($R_1$)
predicted in \cite{branz,raquelx} as a partner of the $X_0(2866)$  state. These states, together with another $J^P=2^+$ partner, are considered molecular states of
$D^*\bar{K}^*$ in $I=0$.
 The reaction proceeds via external emission  and corresponds to
the most favored Cabibbo decay of the $B$. On the other hand the $D^{*+} K^-$  pair couples to $I=0$ and $D^*\bar{K}$ is the only decay channel of the $R_1$ state.  We obtain the  $R_1$ production from  the partner  reaction  $\bar{B}^0 \to   \bar{D}^{*0} D^{*+} K^{*-}$. The
 $D^{*+} K^{*-}$ state propagates and forms the $R_1$ resonance, which later decays to  $D^{*+} K^{-}$. We relate the  mass distribution for the production of $R_1$ with the non resonant background, making a justified assumption which reproduces very well the ratio of signal to
 background in the case of the $X_0(2866)$  production in the $B^- \to D^- D^+ K^+$ reaction. With this scenario we find a ratio of signal to
 background for the $R_1$ production in the $\bar{B}^0 \to \bar{D}^{*0} D^{*+}   K^- $ reaction of a factor of about $7$ at the peak of the
 resonance, and a factor  $0.4$ for the integrated cross  sections. With this  we obtain a branching fraction of about $4 \times 10^{-3}$ for
 the production of the $R_1$  in the $\bar{B}^0 \to \bar{D}^{*0} D^{*+}   K^- $ reaction,  with  $R_1$ observed in the  $D^{*+} K^-$ invariant
 mass distribution. Both the strong peak over the background and the branching ratio are very large, which make very appealing the search of this state obtained
 within the molecular picture for the $X_0(2866)$   and its  partners. Certainly, the observation of this state would be a step forward  to our better understanding the nature of the exotic meson states.

\section*{ACKNOWLEDGEMENT}
This work is partly supported by the National Natural Science Foundation of China under Grants Nos. 12175066, 11975009, 12147219.
RM acknowledges support from the CIDEGENT program of the Generalitat Valenciana with Ref. CIDEGENT/2019/015 and from the Spanish national grants PID2019-106080GB-C21 and PID2020-112777GB-I00.
This work is also partly supported by the Spanish Ministerio de Economia y Competitividad (MINECO) and European FEDER funds
under Contracts No. FIS2017-84038-C2-1-P B, PID2020-112777GB-I00, and by Generalitat Valenciana under contract PROMETEO/2020/023.
This project has received funding from the European Union Horizon 2020 research and innovation programme under
the program H2020-INFRAIA-2018-1, grant agreement No. 824093 of the STRONG-2020 project.

\end{document}